\documentclass[twocolumn,showpacs,preprintnumbers,amsmath,amssymb]{revtex4-1}

\usepackage[english]{babel}

\usepackage{graphicx}
\usepackage{subfigure}
\usepackage{amssymb}
\usepackage{amsfonts}
\usepackage{amsmath}
\usepackage{bm}
\usepackage{latexsym}
\usepackage{verbatim}
\usepackage{epsfig}
\usepackage{float}

\def\pra#1{{ Phys.\ Rev. A\/} {\bf#1}}
\def\prl#1{{ Phys.\ Rev. Lett\/} {\bf#1}}

\def\epl#1{{ Europhys.\ Lett.\/} {\bf#1}}
\def\epjd#1{{ Eur.\ Phys. J. D\/} {\bf#1}}

\def\nat#1{{ Nature Physics} {\bf#1}}

\begin{document}

\title{Crossover Between Non-Markovian and Markovian Dynamics Induced by a Hierarchical Environment}

\author{Tiantian Ma$^1$\footnote{Email address: tma@stevens.edu}, Yusui Chen$^1$, Tian Chen$^{1,2}$, Samuel R. Hedemann$^1$ and Ting Yu$^1$\footnote{Email address:ting.yu@stevens.edu}}

\affiliation{$^{1}$Center for Controlled Quantum Systems  and  Department of Physics and Engineering Physics, Stevens Institute of Technology, Hoboken, New Jersey  07030, USA \\ $^{2}$State Key Laboratory of Low Dimensional Quantum Physics, Department of Physics, Tsinghua University, Beijing 100084, People's Republic of China }

\begin{abstract}
Non-Markovian evolution of an open quantum system can be induced by the memory effects of a reservoir. Although a reservoir with stronger memory effects may seem like it should cause stronger non-Markovian effects on the system of interest, this seemingly intuitive thinking may not always be correct. We illustrate this by investigating a qubit (a two-level atom) that is coupled  to a hierarchical environment, which contains a single-mode cavity and a reservoir consisting of infinite numbers of modes. We show how the non-Markovian character of the system is influenced by the coupling strength between the qubit and cavity and the correlation time of the reservoir. In particular, we found a new phenomenon whereby the qubit Markovian and non-Markovian transition exhibits a anomalous pattern in a parameter space depicted by the coupling strength and the
correlation time of the reservoir.
\end{abstract}

\maketitle

\section{Introduction}

The Markovian approximation  is important and helpful when one is dealing with an open quantum system \cite{Gardiner_QuantumNoise}. This approximation is made by assuming that the correlation function of the reservoir decays much faster than the characteristic time scale of the evolution of the system of interest so that it can be taken as a delta function, and the correlation time, also called ``memory time," is zero.  Under this assumption, a reservoir is sometimes considered Markovian. One advantage of this approximation is that, \textit{in most cases}, the dynamics of the system will be a Markovian process and can be described by a standard Markovian master equation.

However, it has been shown that the Markovian approximation fails in many situations \cite{Mogilevtsev_2008(PRL)_Experimental NM,Galland_2008(PRL)_Non-Markovian Decoherence,Madsen_2010(PRL)_Observation of NM,Tang_2012(EPL)_Measuring non-Markovianity}. One consequence of the breakdown of this approximation is that the evolution of the system becomes non-Markovian rather than Markovian. Thus, the topic of non-Markovian quantum dynamics has recently been studied intensively \cite{BLHu,TingYu_2004(PRA)_QSD,Breuer_2010(PRA)_Exact NM solution,ZhangWeiMin_2012(PRL)_General NM Dynamics,Wolf} and is catching more and more eyes.

To quantify the non-Markovian character of an open system's dynamics, several measures of non-Markovianity (NM) have been proposed \cite{Breuer_2009(PRL)_Definition of NM,Rivas_2010(PRL)_Definition of NM,Luo_2012(PRA)_Def_Of_NM}. With the help of these measures, one can claim that an evolution is non-Markovian if a non-zero NM is detected. These measures have been applied to many models to investigate their non-Markovian features \cite{Breue_2010(PRA)_Cal of NM,ZYXu_2010(PRA)_NM of qubit,BinShao_2011(PRA),TTXu_2012(EPJD)_NM without RWA,Haikka_2013(PRA)_Dephasing NM with Ohmic Spec,Addis_2013(PRA)_TwoQubit NM,Fanchini_2013(PRA)}. Furthermore, a demonstration of control over the transition from Markovian to non-Markovian dynamics has also been experimentally implemented based on these measures \cite{Nature Physics_2011}.

Among these studies, the breakdown of the Markovian approximation plays a crucial role. The breakdown happens if the correlation time is not zero anymore and the reservoir exhibits memory effects. A good example of this is the situation where a single dissipative qubit is coupled to a reservoir with a Lorentzian spectrum \cite{Breuer_2009(PRL)_Definition of NM,Breue_2010(PRA)_Cal of NM}. In this case, the correlation function of the reservoir is an exponential function and the correlation time can be well-defined.  For this model, it has been shown that the dynamics of the qubit is Markovian when the correlation time is very small and non-Markovian when the correlation time is large. Also, a simple monotonic relation between the NM and the correlation time was presented. In some sense, this may not seem surprising since one may intuitively reason that the NM should be larger if the correlation time is larger, allowing the memory effects of the reservoir to be stronger due to the Markovian approximation's failure for large correlation times. However, the transition from non-Markovian to Markovian dynamics is still poorly understood if the environment is not only formed by a bath of free bosons.

The purpose of this paper is to examine the interrelationship between the non-Markovianity and the structured environment. To do so, we will consider a  two-level
system coupled to a composite environment consisting of a single cavity mode and a reservoir with infinite numbers of degrees of freedom.  The model under investigation is simple, yet sophisticated enough to exhibit some interesting features on the non-Markovian and Markov crossover dynamics.  Our major motivation of the present paper is to understand how the structural features of environment affect the non-Markovianity exhibiting the crossover properties between non-Markovian and Markovian regimes.
 It should be pointed out that
non-Markovian dynamics for the same qubit-cavity model has been carefully studied experimentally in \cite{Madsen_2010(PRL)_Observation of NM} without using the non-Markovianity.
For a single reservoir with an Ornstein-Uhlenbeck type of correlation
function, the reservoir correlation time can be easily identified with a single parameter charactering the reservoir decay time.  It should be noticed that such a single
parameter  representing the memory time of the composite environment does not exist in general.  For the composite environment considered in this paper,
it is easy to see that there are several time scales describing the mutual information exchange between two subsystems as well as between the system and its environment.
Hence specifically, we shall investigate in several parameter domains of the cavity-reservoir coupling and the
 memory of the reservoir and see how these parameters affect the system's NM. In particular, we show new crossover properties in non-Markovian-Markov
 transition induced by this hierarchical environment.

We organize the rest of the paper as the following. In Sec.~\ref{Model}, we present a model in which a qubit (the system of interest) is coupled to a hierarchically structured environment consisting of a single-mode cavity dissipatively coupled to a reservoir with a Lorentzian spectrum. In spite of its simplicity, the model provides many useful insights into the non-Markovian dynamics of an open system coupled to a hierarchical environment with the exact solution. The measure of NM is also briefly introduced here. In Sec.~\ref{Discussion} we find that, remarkably, the simple monotonic relation between NM and the correlation time one might intuitively believe may not be valid in this case. Specifically, although the dynamics of the qubit may be non-Markovian when the reservoir has a certain correlation time, we find that it becomes Markovian for \textit{longer} correlation times. Finally, the conclusion is made in Sec~ \ref{Conclusion}.

\section{Model, Solution and Non-Markovianity}\label{Model}

\begin{figure}
\includegraphics[width=3in]{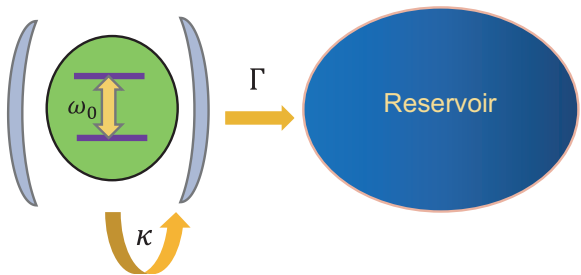}\caption{\label{fig:configuration} (Color online) Configuration of the system plus a hierarchical environment: The qubit of interest is coupled to a single-mode cavity while the cavity is coupled to a reservoir.}
\end{figure}

We consider the following model, the schematic of which is shown in Fig.~\ref{fig:configuration}. The total Hamiltonian can be written as (setting $\hbar=1$)
\begin{equation}
H=H_S+H_C+H_{R}+H_I.
\end{equation}
Here, $H_S=\frac{\omega_s}{2}\sigma_z$ and $H_C=\omega_c a^{\dagger}a$ are the Hamiltonian of the qubit and cavity, $H_{R}=\Sigma_k \omega_k b_k^{\dagger} b_k$ represents the zero-temperature bosonic reservoir, and $H_I$ describes the interactions between the subsystems. If we denote the ground and excited levels of the qubit by $|g\rangle$ and $|e\rangle$ respectively, then $\sigma_z=|e\rangle \langle e|-|g\rangle \langle g|$ is a Pauli matrix. Here $a^\dagger$, $a$ and $b_k^\dagger$ £¬$b$ are the creation and annihilation operators for the cavity and the $k$'s mode of the reservoir, respectively. $\omega_s$ is the transition frequency of the qubit, while $\omega_c$ and $\omega_k$ are the frequencies associated with the cavity and the $k$'s mode of the reservoir, respectively.
For simplicity, we assume $\omega_s=\omega_c=\omega_0$. Then, converting the interaction Hamiltonian $H_I$ to the interaction picture yields
\begin{equation}\small \label{Hamiltonian}
H^{\text{int}}_I=\kappa(\sigma_{+}a+\sigma_{-}a^{\dagger})+\sum_{k}g_{k}(ab_{k}^{\dagger}e^{i\Delta_{k}t}+a^{\dagger}b_{k}e^{-i\Delta_{k}t}),
\end{equation}
where $\Delta_k=\omega_k-\omega_0$, $\kappa$ is the coupling strength between the qubit and cavity, and $g_k$
is the coupling strength between the cavity and the $k$'s mode of the reservoir.
  We suppose that the reservoir has a Lorentzian spectrum
$J(\omega)=\frac{\Gamma}{2\pi}\frac{\lambda^{2}}{(\omega_{0}-\omega)^{2}+\lambda^{2}}$. Then the correlation function of the reservoir is $\alpha(t,s)=\frac{\Gamma \lambda}{2}e^{-\lambda|t-s|}$. Thus $\tau=\lambda^{-1}$ represents the correlation time or memory time. When $\lambda$ goes to infinity, the reservoir converges to a memoryless reservoir without memory effects. For simplicity, we assume that the total environment including both the cavity and reservoir is initially in the vacuum state.
The advantage of this assumption is that the model can be easily solved analytically without losing the features of the physics in which we are interested.

Given these conditions, the cavity stays at the ground level initially and there is always only up to one excitation in the total system. Then the total state can be generally written as \cite{Breuer_OpenQuantumSystem}
\begin{eqnarray}
|\phi(t)\rangle &=& C(t)|g,0,0_{k}\rangle+A(t)|e,0,0_{k}\rangle \nonumber \\
                & \; &+ B(t)|g,1,0_{k}\rangle+\sum_{k}C_{k}(t)|g,0,1_{k}\rangle,\label{eq:total state}
\end{eqnarray}
where $|0\rangle$ and $|1\rangle$ are the vacuum and single-photon states of the cavity, while $|0_{k}\rangle$ represents no excitation in the reservoir, and
$|1_{k}\rangle$ means that there is one excitation in the $k$-th mode
of the reservoir. The dynamics of the qubit can be obtained exactly by partial-tracing both the cavity and reservoir, yielding $\rho=\textrm{Tr}_{C,R}[|\phi(t)\rangle \langle \phi(t)|]$, which has matrix elements (see Appendix~\ref{evolution of qubit})
\begin{eqnarray}
\rho_{ee}(t)  =  \rho_{ee}(0)|G(t)|^{2}, \quad \rho_{eg}(t)  =  \rho_{eg}(0)G(t).\label{eq:Evolution}
\end{eqnarray}
Here, the function $G(t)$ satisfies
\begin{eqnarray}
G(t)  =  L^{-1}[\mathcal{G}(p)],\quad \mathcal{G}(p)  =  \frac{p+\frac{\Gamma\lambda}{2(p+\lambda)}}{p^{2}+\kappa^{2}+\frac{p \Gamma \lambda}{2(p+\lambda)}},\label{eq:Gt}
\end{eqnarray}
where $L^{-1}$ is the inverse Laplace transform. Thus, $G(t)$ is determined analytically for each given set of parameters $\kappa$, $\lambda$, $\Gamma$, with the initial condition $G(0)=1$.

A Markovian evolution can always be represented by a dynamical semigroup of completely positive and trace-preserving (CPT) maps. These properties guarantee the contractiveness of the trace distance (to be defined below) between any fixed pair of initial states $\rho_{1}(0)$ and $\rho_{2}(0)$, which means that a Markovian evolution can never increase the trace distance, it can only decrease it or leave it unchanged. The decrease of trace distance indicates the reduction of distinguishability between the two states. This could be interpreted as an outflow of information from the system to the environment. A violation of this contractive condition is understood as a backflow of information into the system of interest. Based on this concept, a measure of NM can be defined as in \cite{Breuer_2009(PRL)_Definition of NM} by
\begin{equation}
%
\mathcal{N}=\max_{\rho_{1}(0),\rho_{2}(0)}\int_{\sigma>0}dt\sigma(t,\rho_{1}(0),\rho_{2}(0)).\label{eq:non-markovianity_BLP}
\end{equation}
Here, $\sigma(t,\rho_{1}(0),\rho_{2}(0))=\frac{d}{dt}D(\rho_{1}(t),\rho_{2}(t))$ is the rate of change of the trace distance, which is defined as
\begin{equation}
D(\rho_{1}(t),\rho_{2}(t))=\frac{1}{2}\text{Tr}|\rho_{1}(t)-\rho_{2}(t)|,\label{eq:trace distance}
\end{equation}
where $|A|=\sqrt{A^{\dag}A}$.  Thus, $\mathcal{N}$ represents the total increase of distinguishability
over the whole time evolution, i.e., the total amount of information
flowing back to the system of interest. Under this measure, an evolution is non-Markovian if and only if (iff) $\mathcal{N}>0$.
This is also equivalent to saying that an evolution is Markovian if and only if the trace distance of any two initial states decreases monotonically.

In our case, for the evolution in Eq.~(\ref{eq:Evolution}), a monotonically decreasing function $|G(t)|$ is also a necessary and sufficient condition that
the evolution is Markovian \cite{Breue_2010(PRA)_Cal of NM}.
%
Explicitly, given our system's evolution as described by Eq.~(\ref{eq:Evolution}), the trace distance is
\begin{equation}
D(\rho_{1}(t),\rho_{2}(t))=|G(t)|\sqrt{|G(t)|^{2}(\Delta a)^{2}+|\Delta b|^{2}},\label{eq: trace distance_our model}
\end{equation}
where $G(t)$ is given in Eq.~(\ref{eq:Gt}), and $\Delta a=\langle e|\rho_{1}(0)|e\rangle-\langle e|\rho_{2}(0)|e\rangle$,
$\Delta b=\langle e|\rho_{1}(0)|g\rangle-\langle e|\rho_{2}(0)|g\rangle$.
Though optimization is technically needed in Eq.~(\ref{eq:non-markovianity_BLP}), it is not difficult to
see that the detection of NM will be recognized with a non-monotonic
function $|G(t)|$, if one notices that the trace distance $D(t)=D(\rho_{1}(t),\rho_{2}(t))$ in Eq.~(\ref{eq: trace distance_our model})
shares the same monotonicity with $|G(t)|$. More interestingly, if an evolution follows Eq.~(\ref{eq:Evolution}), then the monotonicity of $D(t)$ does \textit{not} depend on the choice of initial states. Thus, the maximization can be removed without affecting the sensitivity of $\mathcal{N}$ for detecting
the NM \cite{Zeng_2011(PRA)_NM equivalence}. Nevertheless, the optimized pair of initial states
we found through numerical simulation is $\rho_{1}=|+\rangle\langle +|$ and $\rho_{2}=|-\rangle\langle -|$, where $|\pm\rangle =\frac{1}{\sqrt{2}}(|e\rangle\pm|g\rangle)$,
which has also been proven theoretically \cite{ZYXu_2010(PRA)_NM of qubit,BinShao_2011(PRA)}.

Particularly in this paper, we numerically integrate Eq. (\ref{eq:non-markovianity_BLP}), with the help of Eq. (\ref{eq: trace distance_our model}), to compute the NM, while the two initial states are taken as $\rho_{1}=|+\rangle\langle +|$ and $\rho_{2}=|-\rangle\langle -|$. Though the values of the NM are obtained by the numerical method, we emphasize that the detection of NM can be done by showing the monotonicity of $D(t)$ analytically whenever the explicit model parameters are given. Our conclusion is not unaffected by the possible numerical errors.

\section{Discussion}\label{Discussion}

\begin{figure}
\includegraphics[width=3.3in]{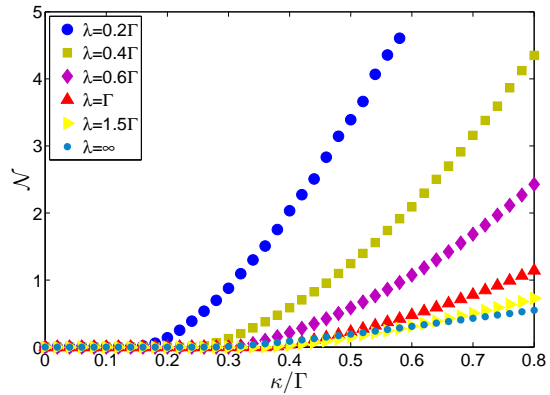}\caption{\label{fig:NM-vs-kappa with varying lambda} (Color online) The change of the NM with respect to $\kappa$ for different $\lambda$. From top to bottom, the $\lambda$ changes from $0.2\Gamma$ to $\infty$.}
\end{figure}

In the following, we discuss how the two parameters $\kappa$ and especially $\lambda$, or the correlation time, influence the NM of the qubit while $\Gamma$ is constant.
First, we focus on $\kappa$. The variation of NM with respect to $\kappa$ for different $\lambda$ is plotted
in Fig.~\ref{fig:NM-vs-kappa with varying lambda}.
For each line (a fixed $\lambda$), the increase of $\kappa$ leads to the growing of NM. An interesting feature here is that a transition from Markovian to non-Markovian dynamics is observed for each line. This fact will also be verified in later discussions.  The speed that the information flowing out of the qubit is very low when $\kappa$ is small, while the evolution of the environment itself is in a very fast pace when $\lambda$ and $\Gamma$ is large. A relatively small $\kappa$ with respect to $\lambda$ and $\Gamma$ indicates that the qubit is losing information at a far slower rate than the environment is evolving, so that the backflow of information cannot happen and the environment is not appreciably interrupted. Thus the phenomenon of transition can only arise from the fact that the coupling strength $\kappa$ becomes so strong that the qubit has disturbed the environment, thereby undermining the foundation of the Markovian approximation, which eventually results in the appearance of information backflow to the qubit.

It is worth noting the situation where the reservoir is memoryless ($\lambda \rightarrow \infty$). In this case, the presence of the cavity is fully responsible for the non-Markovian character. Also, the solution of $G(t)$ in Eq.~(\ref{eq:Gt}) is,
\begin{eqnarray}
G(t) = e^{-\frac{\Gamma t}{4}}[\frac{\Gamma}{a}\sinh(\frac{at}{4})+\cosh(\frac{at}{4})],
\end{eqnarray}
where $a = \sqrt{\Gamma^2-16\kappa^{2}}$. This formally reproduces the results in \cite{Breuer_2010(PRA)_Exact NM solution,Breue_2010(PRA)_Cal of NM}
except for a difference in the scale of parameters. This coincidence stems from the fact that the dynamics of a single qubit coupled to a vacuum reservoir with a Lorentzian spectrum could be simulated by a pseudomode approach with a memoryless reservoir \cite{Garraway_2009PRA}. Two distinct dynamical regimes \cite{Breue_2010(PRA)_Cal of NM} are identified by a threshold $\kappa_{T}=\frac{\Gamma}{4}$. In the weak coupling regime where $\kappa<\kappa_{T}$, the evolution is Markovian and $G(t)$ decreases monotonically. In the strong coupling regime where $\kappa>\kappa_{T}$, the evolution is non-Markovian and $G(t)$ oscillates
between positive and negative values.

\begin{figure}
\includegraphics[width=3.3in]{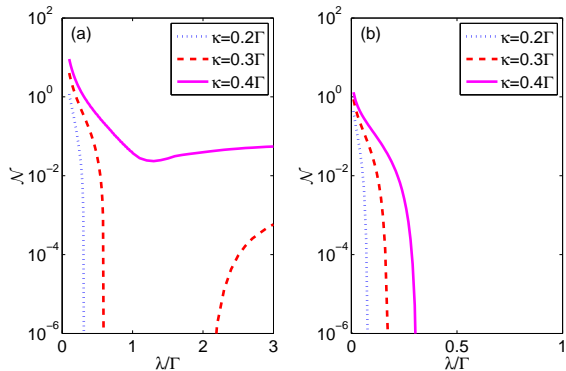}\caption{\label{fig:NM-vs-lambda_comparison} (Color online) The change of NM with respect to $\lambda$ for (a) the model where a qubit is coupled to a hierarchical environment consisting of one cavity and one reservoir with the memory time of $\tau=\lambda^{-1}$, as shown in Fig.~\ref{fig:configuration} and (b) a model where a qubit is coupled to a reservoir with coupling strength $\kappa$ and memory time $\tau =\lambda^{-1}$ of the reservoir. It is worth mentioning that the dashed and dotted lines in (a) and all three lines in (b) decrease exactly to zero though it is not shown in this figure.}
\end{figure}

Now we focus on $\lambda$. Recall that $\tau=\lambda^{-1}$ is the correlation time of the reservoir. When $\lambda$ becomes finite and keeps decreasing, the Markovian approximation of the reservoir fails and one might expect the memory effects of the reservoir to enhance the amount of information backflow, and hence to increase the NM, as well. This would be true if one were considering a model where the qubit is directly connected to a reservoir without the cavity and $\kappa$ is the coupling strength between them, as shown in Fig.~\ref{fig:NM-vs-lambda_comparison}(b).
There, we see a simple monotonic relation between $\lambda$ and NM; a decreasing correlation time ($\lambda$ is moving towards the right) results in a lower value of NM.

However, this relationship may not be universally true. When we consider our hierarchical environment model, $\lambda$ and the NM exhibit non-monotonic relations when $\kappa=0.3\Gamma$ and $\kappa=0.4\Gamma$, as shown in Fig.~\ref{fig:NM-vs-lambda_comparison}(a). The particularly astonishing phenomenon
is that when $\kappa=0.3\Gamma$, the NM drops to zero first and later revives as the parameter $\lambda$ continues to
grow. This revival is due to the fact that $\kappa=0.3\Gamma$ is larger than the threshold $\kappa_{T}(\lambda\rightarrow\infty)=\frac{\Gamma}{4}$. Therefore, when $\kappa=0.3\Gamma$, the evolution of the qubit well eventually become non-Markovian if $\lambda$
is approaching $\infty$ (as the correlation time $\tau\to 0$)!

Thus, the surprising message is that a stronger memory effect of the reservoir may \textit{not} always be helpful in enhancing the NM of the system, due to the presence of the cavity. In fact, because the reservoir is only a part of the environment now, an integrated consideration including both the cavity and the reservoir is needed to determine the non-Markovian character of the qubit of interest. An increase of memory effects from the reservoir alone is not sufficient to estimate the change of NM.

\begin{figure}
\includegraphics[width=3.3in]{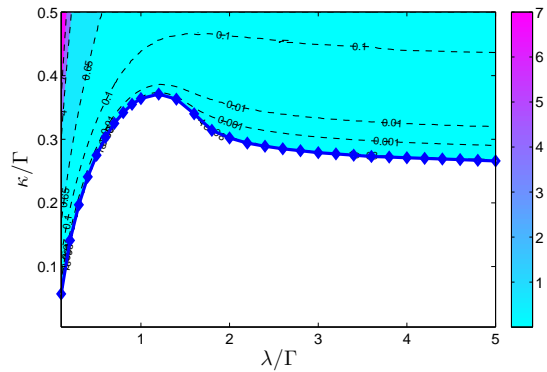}\caption{\label{fig:NM-vs-kappa-lambda_threshold} (Color online) The NM of the qubit for different $\kappa$ and $\lambda$. The Non-Markovian regime is colored while the Markovian regime is white. The dashed black lines are the contour lines of the NM. The diamond blue line is the curve of the threshold $\kappa_T(\lambda)$.}
\end{figure}

To comprehensively explain how our modulation of the environment affects the NM of the qubit, Fig.~\ref{fig:NM-vs-kappa-lambda_threshold} shows how the NM changes with respect to $\kappa$ and $\lambda$.
It is shown that a non-Markovian
threshold $\kappa_{T}(\lambda)$ exists for every given $\lambda$. Thus, the transition from Markovian to non-Markovian dynamics always exists for whatever value $\lambda$ takes, which verifies the statement we made before.

Two interesting regimes are identified clearly in Fig.~\ref{fig:NM-vs-kappa-lambda_threshold}: the white Markovian regime is below the threshold $\kappa_{T}(\lambda)$, and the non-Markovian regime is above $\kappa_{T}(\lambda)$ and is colored.
However, the pattern of $\kappa_{T}(\lambda)$
is rather interesting, shown as the diamond line in Fig.~\ref{fig:NM-vs-kappa-lambda_threshold}.
The curve of the threshold is not
a monotonic function of $\lambda$. The threshold $\kappa_{T}(\lambda)$ increases
as $\lambda$ increases when $\lambda$ is small, which is reasonable
since the memory time of the reservoir is shorter and therefore a larger $\kappa$ is necessary to make a non-Markovian evolution. Nevertheless, the curve
is bent down as $\lambda$ continues to increase and then eventually
approaches to $\frac{\Gamma}{4}$ which is the limit in the memoryless reservoir
case. The overall message here agrees with the statement we made before: the NM does not necessarily decrease as the correlation time of the reservoir decreases. The
non-Markovian dynamics of the qubit is determined by a delicate balance between the two major parameters $\lambda$, and $\kappa$.  This is the major result of this
paper.

Finally, to further demonstrate our result, we directly investigate
the trace distance $D(t)$ given in Eq.~(\ref{eq: trace distance_our model}).
\begin{figure}
\includegraphics[width=3.3in]{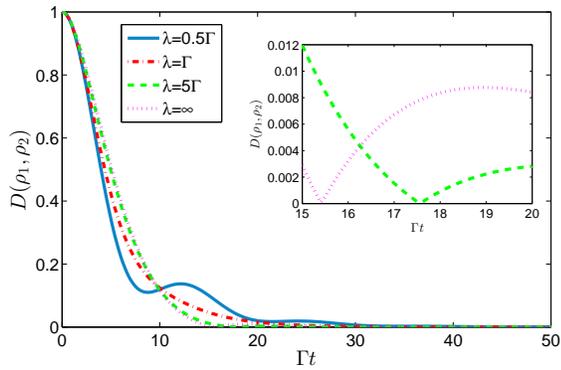}\caption{\label{fig:,-Evolution-of-trace-distance} (Color online) The evolution of trace distance $D(t)$ when $\kappa=0.3\Gamma$ and the pair of initial states is $\rho_{1}(0)=|+\rangle\langle +|$ and  $\rho_{2}(0)=|-\rangle\langle -|$, where $|\pm\rangle=\frac{1}{\sqrt{2}}(|e\rangle\pm|g\rangle)$.}
\end{figure}
Figure~\ref{fig:,-Evolution-of-trace-distance} shows its evolution
when $\kappa=0.3\Gamma$.
If $\lambda=0.5\Gamma$, we are in the non-Markovian regime. $D(t)$ is not monotonic and the evolution is non-Markovian. When $\lambda$ increases to $\Gamma$, we arrive at the Markovian regime. $D(t)$ becomes monotonic and the evolution becomes Markovian. However, $D(t)$ becomes non-monotonic again when $\lambda$ continues to increase as we are falling to the non-Markovian regime once more.

A notable point is that even in the non-Markovian regime, $D(t)$ exhibits different patterns for different $\lambda$s. When $\lambda=0.5\Gamma$,
the curve of $D(t)$ is bumpy, but gradually approaches to zero.
However, for the cases $\lambda=5\Gamma$ and $\lambda=\infty$, $D(t)$
keeps hitting the zero line and then bounces back, as seen in the inset of Fig.~\ref{fig:,-Evolution-of-trace-distance}. These zero-points mean
the two states $\rho_{1}$ and $\rho_{2}$ are totally indistinguishable
at those time-points and correspond to the points where $G(t)=0$. From Eq.~(\ref{eq:Evolution}), one can tell that the qubit actually evolves into its ground state at these zero points, and hence loses all the information. The qubit is supposed to stop evolving after this point without recapturing the lost information under a typical Markovian evolution. Thus, the bounce of $D(t)$ from the the inset of Fig.~\ref{fig:,-Evolution-of-trace-distance} serves as a remarkably non-Markovian feature, meaning that the information could flow back into the qubit even if it has been completely leaked into the environment, which would never happen in a Markovian evolution.

\section{Conclusion}\label{Conclusion}

In summary, we studied a qubit that is coupled to a hierarchically structured
environment consisting of a cavity and a reservoir. We investigated how the qubit-cavity coupling strength and the reservoir's memory time
affect the non-Markovian character of the qubit. We found that a threshold $\kappa_{T}(\lambda)$
exists for an arbitrarily given $\lambda$, separating the Markovian and non-Markovian
regimes of the parameter space. Surprisingly, $\kappa_{T}(\lambda)$ is a non-monotonic
function of $\lambda$ and a longer correlation time of the reservoir does not necessarily result in a larger value of NM.

Finally, it should be noted that our calculation is based on the measure of the NM proposed in \cite{Breuer_2009(PRL)_Definition of NM}. Several other measures of the NM have been proposed as well \cite{Rivas_2010(PRL)_Definition of NM,Luo_2012(PRA)_Def_Of_NM}. Generally, these measures do not need to agree with each other \cite{Rivas_2011(PRA)_NM are not equivalent}. However, it has been proven that they are equivalent in the sense of detecting the NM for the dynamics in the form of Eq.~(\ref{eq:Evolution}) \cite{Zeng_2011(PRA)_NM equivalence,Luo_2012(PRA)_Def_Of_NM}. Therefore, our conclusion is invariant with respect to the definition of the NM.

\section*{Acknowledgement}
We acknowledge grant support from DOD/AF/AFOSR No. FA9550-12-1-0001.
TY is grateful to Prof. H. S. Goan for the hospitality during his visit to the National Taiwan University. \\

\appendix
\section{Evolution of the Qubit} \label{evolution of qubit}

Plug the state in Eq.~\ref{eq:total state} and the Hamiltonian in Eq.~\ref{Hamiltonian} into the
Schrodinger's Equation $|\dot{\phi}(t)\rangle=-iH_{I}^{int}|\phi(t)\rangle$,
we obtain the following:

\begin{eqnarray}
\dot{A}(t) & = & -i\kappa B(t),\nonumber\\
\dot{B}(t) & = & -i\kappa A(t)-i\sum_{k}g_{k}e^{-i\Delta_{k}t}C_{k}(t)d\tau,\nonumber\\
\dot{C}_{k}(t) & = & -ig_{k}e^{i\Delta_{k}t}B(t),\nonumber\\
C(t) & = & C(0).
\end{eqnarray}

Considering the initial conditions that $B(0)=C_{k}(0)=0$ and the
correlation function $\alpha(t,s)=\sum_{k}|g_{k}|^{2}e^{-i\Delta_{k}(t-s)}=\frac{\Gamma\lambda}{2}e^{-\lambda|t-s|}$,
we have:
\begin{eqnarray}
\dot{A}(t) & = & -i\kappa B(t), \nonumber\\
\dot{B}(t) & = & -i\kappa A(t)-\int_{0}^{t}\alpha(t-\tau)B(\tau)d\tau.
\end{eqnarray}

Taking advantage of the Laplace transform $\mathcal{F}(p)\equiv L[F(t)]=\int_0^\infty F(t)e^{-pt}dt$ leads to
\begin{eqnarray}
p\mathcal{A}(p)-A(0) & = & -i\kappa \mathcal{B}(p), \nonumber\\
p\mathcal{B}(p)-B(0) & = & -i\kappa \mathcal{A}(p)-\frac{\Gamma \lambda}{2(p+\lambda)}\mathcal{B}(p).
\end{eqnarray}
Then we easily achieve $\mathcal{A}(p)=A(0)\mathcal{G}(p)$ and $A(t)=A(0)G(t)$,
where $\mathcal{G}(p)$ and $G(t)$ are given in Eq.~(\ref{eq:Gt}). The state of the qubit of interest
is then given by

\begin{equation}
\rho=\textrm{Tr}_{C,R}[|\phi(t)\rangle\langle\phi(t)|]=\left(\begin{array}{cc}
|A(t)|^{2} & A(t)C(0)^{*}\\
A(t)^{*}C(0) & 1-|A(t)|^{2}
\end{array}\right),
\end{equation}

which satisfies Eq.~(\ref{eq:Evolution}).



\begin{thebibliography}{10}

\bibitem[1]{Gardiner_QuantumNoise}
C. W. Gardiner and P. Zoller, Quantum Noise (Springer Verlag, Berlin, 2000).

\bibitem[2]{Mogilevtsev_2008(PRL)_Experimental NM}
D. Mogilevtsev, A. P. Nisovtsev, S. Kilin, S. B. Cavalcanti, H. S. Brandi, and L. E. Oliveira, \prl{100}, 017401 (2008).

\bibitem[3]{Galland_2008(PRL)_Non-Markovian Decoherence}
C. Galland, A. Hogele, H. E. Tureci, and A. Imamoglu, \prl{101}, 067402 (2008).

\bibitem[4]{Madsen_2010(PRL)_Observation of NM}
K. H. Madsen, S. Ates, T. Lund-Hansen,, A. Loffler, S. Reitzenstein, A. Forchel, and P. Lodahl, \prl{106}, 233601 (2011).

\bibitem[5]{Tang_2012(EPL)_Measuring non-Markovianity}
J.-S. Tang, C.-F. Li, Y.-L. Li, X.-B. Zou, G.-C. Guo, H.-P. Breuer, E.-M. Laine, and J. Piilo, \epl{97}, 10002 (2012).

\bibitem[6]{BLHu} B. L. Hu, J. P. Paz, and Y. Zhang, Phys. Rev. D {\bf 45}, 2843 (1992).

\bibitem[7]{TingYu_2004(PRA)_QSD}
W. T. Strunz and T. Yu, \pra{69}, 052115 (2004).

\bibitem[8]{Breuer_2010(PRA)_Exact NM solution}
B. Vacchini and H.-P. Breuer, \pra{81}, 042103 (2010).

\bibitem[9]{ZhangWeiMin_2012(PRL)_General NM Dynamics}
W.-M. Zhang, P.-Y. Lo, H.-N. Xiong, M. W.-Y. Tu, and F. Nori, \prl{109}, 170402 (2012).

\bibitem[10]{Wolf}
M. M. Wolf, J. Eisert, T. S. Cubitt and J. I. Cirac, \prl{101}, 150402 (2008)

\bibitem[11]{Breuer_2009(PRL)_Definition of NM}
H.-P. Breuer, E.-M. Laine, and J. Piilo, \prl{103}, 210401 (2009).

\bibitem[12]{Rivas_2010(PRL)_Definition of NM}
A. Rivas, S. F. Huelga, and M. B. Plenio, \prl{105}, 050403(2010).

\bibitem[13]{Luo_2012(PRA)_Def_Of_NM}
S. Luo, S. Fu, and H. Song, \pra{86}, 044101 (2012).

\bibitem[14]{Breue_2010(PRA)_Cal of NM}
E.-M. Laine, J. Piilo, and H.-P. Breuer, \pra{81}, 062115 (2010).

\bibitem[15]{ZYXu_2010(PRA)_NM of qubit}
Z. Y. Xu, W. L. Yang, and M. Feng, \pra{81}, 044105 (2010).

\bibitem[16]{BinShao_2011(PRA)}
Z. He, J. Zou, L. Li, and B. Shao, \pra{83}, 012108 (2011).

\bibitem[17]{TTXu_2012(EPJD)_NM without RWA}
H.S. Zeng, N. Tang, Y. P. Zheng, and T. T. Xu, \epjd{66}, 255 (2012).

\bibitem[18]{Haikka_2013(PRA)_Dephasing NM with Ohmic Spec}
P. Haikka, T. H. Johnson, and S. Maniscalco1, \pra{87}, 010103 (2013).

\bibitem[19]{Addis_2013(PRA)_TwoQubit NM}
C. Addis, P. Haikka, S. McEndoo, C. Macchiavello, and S. Maniscalco, \pra{87}, 052109 (2013).

\bibitem[20]{Fanchini_2013(PRA)}
F. F. Fanchini, G. Karpat, L. K. Castelano, and D. Z. Rossatto, \pra{88}, 012105 (2013).

\bibitem[21]{Nature Physics_2011}
B.-H. Liu, L. Li, Y.-F. Huang, C.-F. Li, G.-C. Guo,	 E.-M. Laine, H.-P. Breuer, and J. Piilo, \nat{7}, 931-934 (2011).

\bibitem[22]{Breuer_OpenQuantumSystem}
H. P. Breuer and F. Petruccione, The Theory of Open Quantum
Systems (Oxford University Press, Oxford, 2002).

\bibitem[23]{JingJun_2010(PRL)}
J. Jing and T. Yu, \prl{105}, 240403 (2010).

\bibitem[24]{XinyuZhao_2011(PRA)}
X. Zhao, J. Jing, B. Corn, and T. Yu, \pra{84}, 032101 (2011).

\bibitem[25]{Garraway_2009PRA}
L.  Mazzola, S. Maniscalco, J. Piilo, K.-A. Suominen, and B. M. Garraway, \pra{80}, 012104 (2009).

\bibitem[26]{Rivas_2011(PRA)_NM are not equivalent}
D. Chruscinski, A. Kossakowski, and A. Rivas, \pra{83}, 052128 (2011).

\bibitem[27]{Zeng_2011(PRA)_NM equivalence}
H.-S. Zeng, N. Tang, Y.-P. Zheng, and G.-Y. Wang, \pra{84}, 032118 (2011).


\end{thebibliography}
\end{document}